\documentclass[twocolumn,showpacs,preprintnumbers,amsmath,amssymb]{revtex4}

\usepackage{graphicx}
\usepackage{dcolumn}
\usepackage{bm}

\begin{document}


\title{Superconducting Topological Fluids \\ in Josephson Junction Arrays}

\author{M. Cristina Diamantini}
\email{cristina.diamantinitrugenberger@cern.ch}
\affiliation{%
Dipartimento di Fisica and Sezione INFN, University of Perugia, via
A. Pascoli, I-06100 Perugia, Italy
}%

\author{Pasquale Sodano}
\altaffiliation[Present Address: ]{Progetto Lagrange, Fondazione CRT
and Fondazione ISI,  Dipartimento di  Fisica, Politecnico di Torino,
Corso Duca degli Abruzzi 24,
I-10129 Torino, Italy}
\email{pasquale.sodano@pg.infn.it}
\affiliation{%
Dipartimento di Fisica and Sezione INFN, University of Perugia, via
A. Pascoli, I-06100 Perugia, Italy
}%

\author{Carlo A. Trugenberger}
\email{ca.trugenberger@InfoCodex.com}
\affiliation{%
InfoCodex S.A., av. Louis-Casai 18, CH-1209 Geneva, Switzerland\\
Theory Division, CERN, CH-1211 Geneva 23, Switzerland
}%


\date{\today}

\begin{abstract}
We argue that frustrated Josephson junction arrays may support a
topologically ordered superconducting ground state, characterized by
a non-trivial ground state degeneracy on the torus. This
superconducting quantum fluid provides an explicit example of a
system in which superconductivity arises from a topological
mechanism rather than from the usual Landau-Ginzburg mechanism.

\end{abstract}

\maketitle

It is by now widely believed that quantum phase transitions describe
changes in the entanglement pattern of the complex-valued ground
state wave function and that the universality classes of these
quantum ground states define the corresponding quantum orders
\cite{Wen1}. When there is a gap in the spectrum the quantum ordered
ground state is called topologically ordered \cite{Wen2}; its
hallmark is a ground state degeneracy depending only on  the
topology of the underlying space.

The best known example of topological order is given by Laughlin's
quantum incompressible fluids \cite{Laughlin2} describing the ground
states responsible for the quantum Hall effect \cite{GirvinPrange}.
Soon after Laughlin's discovery, it was conjectured that an
analogous mechanism could enable superconductivity \cite{Wilczek} in
high- temperature or granular superconductors, although the original
anyon superconductivity mechanism has now been ruled out
experimentally, due to the observed parity and time reversal
invariance of the ground states of the relevant physical systems.

Josephson junction arrays (JJA) have been regarded by several
authors \cite{ds}, \cite{as}, \cite{dst}, \cite{joffe} as
controllable devices, which may exhibit topological order. Planar
arrays display a characteristic insulator-superconductor quantum
phase transition at $T=0$ \cite{jja}. In \cite{dst} we pointed out
that two-dimensional JJAs may be mapped onto an Abelian gauge theory
with Chern-Simons term and evidenced how the topological gauge
theory, which naturally allows for the appearance of topological
order, together with duality may be useful to describe the phase
diagram of these devices.

The gauge theory formulation of JJAs \cite{dst} clearly evidences
that the superconducting ground state is a P- and T- invariant
generalization of Laughlin's incompressible quantum fluid. The
simplest example of a topological fluid \cite{fwz} is a ground-state
described by a low-energy effective action given solely by the
topological Chern-Simons term \cite{chernsimons} $S = k/4\pi\ \int
d^3x \ A_{\mu} \epsilon^{\mu \nu \alpha}
\partial_{\nu} A_{\alpha}$ for a compact $U(1)$ gauge field
$A_{\mu}$ whose dual field strength $F^{\mu} = \epsilon^{\mu \nu
\alpha} \partial_{\nu} A_{\alpha}$ yields the conserved matter
current. In this case the degeneracy of the ground state on a
manifold of genus $g$ will be $k^g$ (or $(k_1 k_2)^g$ if $k =
k_1/k_2$ is a  rational number): for planar unfrustrated JJAs one
finds that the topological fluid is described by two $k=1$
Chern-Simons gauge fields of opposite chirality and, thus, there is
no degeneracy of the ground state \cite{dst}, \cite{prl}.

In this letter we argue that frustrated JJAs may support a
topologically ordered ground state described by a pertinent
superconducting quantum fluid, thus providing an interesting and
explicit example of a system in which superconductivity arises from
the topological mechanism proposed in \cite{prl} rather than from
the usual Landau-Ginzburg mechanism. In presence of $n_q$ offset
charge quanta per site and $n_{\phi }$ external magnetic flux quanta
per plaquette in specific ratios, Josephson junction arrays might
support incompressible quantum fluid \cite{Lau} \ phases
corresponding to purely two-dimensional quantum Hall phases for
either charges \cite{odi} \ or vortices \cite{choi}. In this paper
we shall show that, if quantum Hall phases for charges or vortices
are realized, then JJA naturally support a topologically ordered
ground state and a phase in which they behave as a topological
superconductor \cite{prl}; there is, in fact, a renormalization of
the Chern-Simons coefficient yielding a non-trivial ground state
degeneracy on the torus (and in general on manifolds with
non-trivial topology).

We shall consider JJAs fabricated on a square planar lattice of
spacing $l = 1$ made of superconducting islands with nearest neighbours
Josephson couplings of strength $E_J$ \cite{jja}. Each island has a
capacitance $C_0$ to the ground; moreover, there are nearest
neighbours capacitances $C$. To implement a torus topology we impose
doubly periodic conditions at the boundary of the square lattice.

In \cite{dst} we have shown that the zero-temperature partition
function of JJAs may be written in terms of two effective gauge
fields $A_{\mu}$ (vector) and $B_{\mu}$ (pseudovector). In the low energy limit the partition
function is
\begin{eqnarray}
Z &&= \sum_{\{ Q_0 \} \atop \{ M_0 \} } \
\int {\cal D}A_{\mu } \int {\cal D}B_{\mu } \ {\rm exp}(-S)\ ,
\nonumber \\
S &&= \int dt \sum_{\bf x} -i{ 1\over 2\pi }\ A_{\mu }K_{\mu \nu }B_{\nu } + i
 A_0 Q_0 + i  B_0 M_0\ .
\label{ac}
\end{eqnarray}
This form of the partition function holds true also with toroidal
boundary conditions. The gauge fields embody the original degrees of
freedom through their dual field strengths $q_{\mu} \propto K_{\mu \nu}
B_{\nu}$ and $\phi_{\mu} \propto K_{\mu \nu} A_{\nu}$ representing,
respectively, the conserved charge (vector) current and the
conserved vortex (axial) current. $K_{\mu \nu}$ is the lattice Chern-Simons term
\cite{latticecs}, defined by $K_{00} = 0$, $K_{0i} = -\epsilon_{ij}
d_j$, $K_{i0} = S_i \epsilon_{ij} d_j$ and $K_{ij} = -S_i
\epsilon_{ij} \partial_0$, in terms of forward (backward) shift and
difference operators $S_i$ ($\hat S_i$) and $d_i$ ($\hat d_i$). Its
conjugate $\hat K_{\mu \nu}$ is defined by $\hat K_{00} = 0$, $\hat
K_{0i} = - \hat S_i \epsilon_{ij} \hat d_j$, $\hat K_{i0} =
\epsilon_{ij} \hat d_j$ and $\hat K_{ij} = - \hat S_j \epsilon_{ij}
\partial_0$. The two Chern-Simons kernels $K_{\mu \nu}$ and $\hat
K_{\mu \nu}$ are interchanged upon integration (summation) by parts
on the lattice.

The topological excitations are described by the integer-valued
fields $Q_0$ and $M_0$ and represent unit charges and vortices
rendering the gauge field components $A_0$ and $B_0$ integers via
the Poisson summation formula; their fluctuations determine the
phase diagram \cite{dst}. In the classical limit
the magnetic excitations are dilute and the charge excitations condense
rendering the system a superconductor: vortex confinement amounts
here to the Meissner effect. In the quantum limit, the magnetic
excitations condense while the charged ones become dilute: the
system exhibits insulating behavior due to vortex superconductivity
accompanied by a charge Meissner effect.

 By rewriting the topological excitations as the
curl of an integer-valued  field
\begin{eqnarray}Q_{0 } &\equiv K_{0 i }Y_{i }\ ,
\qquad \qquad Y_{i }\in Z\ ,\nonumber \\
M_{0 } &\equiv \hat K_{0 i } X_{i } \ ,\qquad \qquad
X_{i } \in Z\ ,
\label{fre}
\end{eqnarray}
we get the mixed Chern-Simons term as follows:

\begin{eqnarray}
Z &= \sum_{\{ X_i, Y_i \}} \
\int {\cal D}A_{\mu } \int {\cal D}B_{\mu } \ {\rm exp}\ (-S)\ ,
\nonumber \\
S &= -{1\over 2\pi }i \int dt \sum_{\bf x} A_0 K_{0i}
\left( B_i - 2 \pi Y_i  \right) \nonumber \\ &+ B_0 \hat K_{0i}
\left( A_i - 2 \pi X_i  \right)
+ A_i K_{ij} B_j \ . \label{ai}
\end{eqnarray}
From (\ref{ai}) one sees that the gauge field components $A_i$ and
$B_i$ are angular variables due to their invariance under
time-independent integer shifts. Such shifts do not affect the last
term in the action, which contains a time derivative, and may be
reabsorbed in the topological excitations $X_i$ and $Y_i$, leaving
also the first term of the action invariant. The low energy theory
is thus compact.

In analogy with the conventional quantum Hall setting one should
expect the charge and vortex transport properties to depend on the
ratios of the offset charges (i.e. the filling fractions)
$(n_q/n_{\phi})$ and $(n_{\phi}/n_q)$, respectively. Due to the
 periodicity of the charge-vortex coupling, however, $n_{\phi }$ ($n_q$)
is defined only modulo an integer as far as charge (vortex)
transport properties are concerned. Using this freedom one may
define effective filling fractions (we shall assume $n_q \ge 0$,
$n_{\phi}\ge 0$ for simplicity) as
\begin{eqnarray}\nu _q &\equiv
{n_q \over {n_{\phi }-\left[ n_{\phi } \right] ^- +
\left[ n_q \right] ^+ }} \ , \qquad \qquad 0\le \nu_q\le 1\ ,\nonumber \\
\nu_{\phi } &\equiv {n_{\phi } \over {n_q-\left[ n_q \right] ^- +\left[
n_{\phi } \right] ^+}} \ ,\qquad \qquad 0\le \nu_{\phi }\le 1 \ ,
\label{eff}
\end{eqnarray}
where $\left[ n_q \right] ^{\pm }$ indicate the smallest (greatest)
integer greater (smaller) than $n_q$. Of course, these effective
filling fractions are always smaller than 1.

In \cite{dst} we assumed the existence of these quantum Hall phases
and discussed them in the framework of the gauge theory
representation of Josephson junction arrays, showing that, depending
on certain parameters of the array there are both a charge quantum
Hall phase and a vortex quantum Hall phase. Here we will concentrate
on the low energy limit of the charge quantum Hall phase and we will
show  that the system has topological order and behaves as a
superconductor when charge condenses.

The pertinent low energy theory is now given by:
\begin{equation} S = \int dt\sum_{\bf x} - {i \over \pi}
A_{\mu }K_{\mu  \nu }B_{\nu } - {i \nu _q
\over \pi }A_{\mu }K_{\mu  \nu }A_{\nu } \ ,
\label{csag}
\end{equation}
with $\nu_q =p/n$. The main difference with (\ref{ac}) \ is the
addition of a  pure Chern-Simons term for the $A_{\mu }$ gauge field
. We have also rescaled the coefficient of the mixed Chern-Simons
coupling by a factor of 2 (compare with (\ref{ac}) ). This factor of
2 is a well-known aspect of Chern-Simons gauge theories \cite{wil}.
Moreover, since in JJAs the charge degrees of freedom are bosons,
the allowed \cite{dst} filling fractions are given by $
\nu_q={p\over n}$ , with $ pn = {\rm even \ integer}$ in accordance
with \cite{read}. As a result, the action (\ref{csag}) may now be
written in terms of two independent gauge fields $A_\mu$ and
$B_\mu^q = B_\mu + \nu_q A_\mu$ yielding:
\begin{equation} S = \int dt\sum_{\bf x} - {i \over \pi}
A_{\mu }K_{\mu  \nu }B_{\nu }^q \ .
\label{csnf}
\end{equation}

In describing JJAs one has to require the periodicity of
charge-vortex couplings; the coupling of the topological excitations
enforcing the periodicity of the mixed Chern-Simons term $A_{\mu
}K_{\mu \nu }B_{\nu }^q$ is then:
\begin{equation}
S = \int dt \sum _x \dots + ip A_{0 }
Q_{0 } + in B_{0 }M_{0 } \ ,
\label{topex}
\end{equation}
that can be rewritten as:
\begin{equation}
S = \sum _x \dots + ilp A_{0 } \left( Q_{0 }+M_{0 } \right) + iln
B^q_{0 }M_{0} \ .\label{ntope}
\end{equation}
Due to the replacement $B_{\mu }\to B_{\mu }^q$, the periodicities
of the two original gauge fields are
\begin{eqnarray}A_{i } &\to A_{i } + \pi n
\ a_{i } \ , \qquad \qquad a_{i } \in Z\ ,\nonumber \\
B_{i } &\to B_{i } + \pi p \ b_{i } \ ,\qquad
\qquad b_{i }\in Z \ ,
\label{newshiq}
\end{eqnarray}
and
\begin{equation}
B^q_{i } \to B^q_{i } + \pi p \ b_{i } \ ,\qquad
\qquad b_{i }\in Z \ .
\label{bper}
\end{equation}
The resulting low energy theory is thus, again, compact.

Using the representation (\ref{fre}), one may rewrite the mixed
Chern-Simons term as
\begin{eqnarray}S = \int dt  \sum_x {\dots } &-i{(pq /2)\over 2\pi}
 \left({2 A_{0 }\over n}
\right) K_{0 i} \left({2 B^q_{i }\over p}
- 2\pi Y_{i } \right) \nonumber \\
 &-i{(pq /2)\over 2\pi}{2 B^q_{0 }\over n}
 K_{0 i } \left({2 A_{i}\over n}
- 2\pi X_{i } \right) \ .
\label{reab}
\end{eqnarray}
In this representation it is clear that the topological excitations render
the charge-vortex coupling  periodic under the shifts
\begin{eqnarray}A^{'}_{i } = {2 A_{i }\over n} &\to A^{'}_{i } + 2\pi
\ a_{i }\ ,\qquad \qquad a_{i } \in Z\ , \nonumber \\
B^{'}_{i } = {2 B^q_{i}\over p} &\to B^{'}_{i } + 2\pi  \ b_{i }\ ,\qquad
\qquad b_{i } \in Z\ .
\label{shi}
\end{eqnarray}
This model corresponds to two Chern-Simons terms with coefficients $
\pm k/4 \pi$ with $k = n p/2$ an integer. It is worth to point out
that, since $B_\nu^q$ does not have a definite parity (is a liner
combination of a vector and a pseudovector) the model is not P- and
T-invariant, as it must be due to the presence  of the Chern-Simons
term for the field $A_\mu$.

The hallmark of topological order is the degeneracy of the ground
state on manifolds with non-trivial topology as shown by Wen
\cite{Wen2}. The torus degeneracy on the lattice of the Chern-Simons
model was computed in \cite{elsem}. For a single Chern-Simons term
this degeneracy is $(k)^g$ where $k$ is the integer coefficient of
the Chern-Simons term, and $g$ the genus of the surface. In our case
this degeneracy is $2 \times (k)^g = 2 \times {n p \over 2}$, since
we have two Chern-Simons terms. This degeneracy is exactly what is
expected for a doubled Chern-Simons model \cite{doubled}, for which
the physical Hilbert space is the direct product of the two Hilbert
spaces of the component models.

We will now demonstrate that the phase where topological excitations $Q_0$
condense while $M_0$ are dilute describes an
effective gauge theory of a superconducting state. The partition function is:
\begin{eqnarray}
Z_{LE} &&= \sum_{\{ Q_0 \} } \
\int {\cal D}A_{\mu } \int {\cal D}B_{\mu }^q \ {\rm exp}\ (-S)\ ,
\nonumber \\
S &&=  \int dt \sum_{\bf x} -{i k \over 2 \pi }\ A_{\mu }K_{\mu \nu }B_{\nu }^q
+ {i k \over 2 \pi } A_0 (2\pi  Q_0)\ .
\label{af}
\end{eqnarray}

 To this end note
first that a unit external charge, represented by an additional term
$ i2\pi a_0 (t, {\bf x}) \delta_{{\bf x}{\bf x_0}}$  is completely
screened by the charge condensate, since it can be absorbed into a
redefinition of the topological excitations $Q_0$. In order to
characterize the superconducting phase we introduce the typical
order parameter namely the 't Hooft loop of length $T$ in the time
direction:
\begin{equation}
L_H \equiv {\rm exp} \left( i \phi {\kappa \over 2 \pi}
\int dt \sum_x \phi _{\mu }B_{\mu }\right) \ ,
\label{thlo}
\end{equation}
 where $\phi_0 (t, {\bf x})$ = $\left( \theta
\left( t+T/2 \right) - \theta \left( t-T/2 \right) \right)
\delta_{{\bf x} {\bf x_1}}$ - $\left( \theta \left( t+T/2 \right) -
\theta \left( t-T/2 \right) \right) \delta_{{\bf x} {\bf x_2}}$ and
$\phi_i (-T/2, {\bf x})$, $\phi_i (T/2, {\bf x})$ are unit links joining
${\bf x_1}$ to ${\bf x_2}$ and ${\bf x_2}$ to ${\bf x_1}$ at fixed
time and vanishing everywhere else. Its vacuum expectation value
$\langle L_H \rangle $ yields the amplitude for creating a separated
vortex-antivortex pair of flux $\phi $, which propagates for a time
$T$ and is then annihilated in the vacuum.

Since we replaced $B_{\mu } \to   B_{\mu }^q$, we may rewrite the 't
Hooft loop as:
\begin{equation}
L_H \equiv {\rm exp} \left( i \phi {\kappa \over 2 \pi}
\int dt \sum_x \left( \phi _{\mu }B_{\mu }^q - {p\over n}  A_\mu \phi_\mu \right)
\right) \ .
\label{nthlo}
\end{equation}

To compute $\langle L_H
\rangle $ one should integrate first over the gauge field $B_{\mu }^q$
to get
\begin{eqnarray}
\langle L_H \rangle &&\propto \sum_{\{ Q_0 \} } \ \int {\cal
D}A_{\mu } \  \delta \left( \hat K_{\mu \nu} A_{\nu} - \phi \phi_{\mu}
\right) \times
\nonumber \\
&& \ {\rm exp} i \phi {\kappa \over 2 \pi}\left( \int dt \sum_{\bf x}
 A_0 (2\pi  Q_0) - {p\over n}  A_\mu \phi_\mu
\right) \ .
\label{ah}
\end{eqnarray}
The sum over $Q_0$ enforces the condition that ${k\over 2 \pi}A_0$ be an integer.
As a consequence, $\hat K_{i 0} A_0 = {2 \pi \over k} n_i = \phi
\phi_i$ with $n_i$ an integer.
We thus have:
\begin{equation}\phi = {2 \pi \over k} q, q  \in Z \ ;
\label{fluq}
\end{equation}
thus, $\langle L_H \rangle $ vanishes for
all fluxes different from an integer multiple of the fundamental
fluxon, which is just the Meissner effect. In the low-energy
effective gauge theory vortex-antivortex pairs are confined by an
infinite force which becomes logarithmic upon including also
higher-order Maxwell terms.

From (\ref{reab}) one has (depending if $p$ is an even integer or
$n$ is an even integer) either that $n$ is the charge unit with
$p/2$ units of charge, or, viceversa, that $p$ is the charge unit
with $n/2$ units of charge. By rewriting (\ref{fluq}) as $\phi = {2
\over pn} 2 \pi n$ one finds the standard flux quantization $\phi =
{2 \pi \over N e} n$ where $e$ is the charge unit and $N$ is the
number of units of charge.

Summarizing, frustrated planar JJAs in the quantum Hall phases
provide an explicit example of both topological order  and of a new
superconducting behavior \cite{prl} analogous to Laughlin's quantum
Hall fluids.

\end{document}